\documentclass[a4paper,11pt]{article}
\usepackage{pos}

\usepackage{wrapfig}

\title{Testing nonstandard neutrino properties}

\author*[a,b]{Manibrata Sen}

\affiliation[a]{Department of Physics, Indian Institute of Technology Bombay,\\ Powai, Mumbai 400076, India}

\affiliation[b]{Max-Planck-Institut für Kernphysik,
\\Saupfercheckweg 1, 69117 Heidelberg, Germany}

\emailAdd{manibrata@iitb.ac.in}

\abstract{Neutrinos provide a compelling avenue to explore physics beyond the Standard Model (SM). This proceeding is a brief summary of a plenary talk given at the 12th Neutrino Oscillation Workshop (NOW2024). We present a discussion on various topics on neutrino non-standard properties, such as the origin of neutrino mass, their decay modes, interaction mechanisms, and a potential connection to dark matter. Constraints and observational results from cosmology, astrophysics, and laboratory experiments are reviewed to illustrate the interplay between neutrino physics and broader questions about the universe. }

\FullConference{12th Neutrino Oscillation Workshop (NOW2024)\\
 2-8, September 2024\\
Otranto, Lecce, Italy\\}

\begin{document}
\maketitle

\section{Introduction}
The discovery of neutrino oscillations has firmly established that neutrinos have mass, a direct indication of the existence of physics beyond the Standard Model (SM)~\cite{ParticleDataGroup:2024cfk}. However, the precise origin and characteristics of this mass remain enigmatic. Furthermore, neutrinos, owing to their minimal interactions with other particles, are uniquely sensitive to new physics. They serve as a testing ground for exploring a variety of extensions of the SM, such as sterile neutrinos, non-standard interactions, lepton number violation, and might play an important role in the dark matter paradigm.

\begin{wrapfigure}{r}{0.4\textwidth}
    \includegraphics[width=0.4\textwidth]{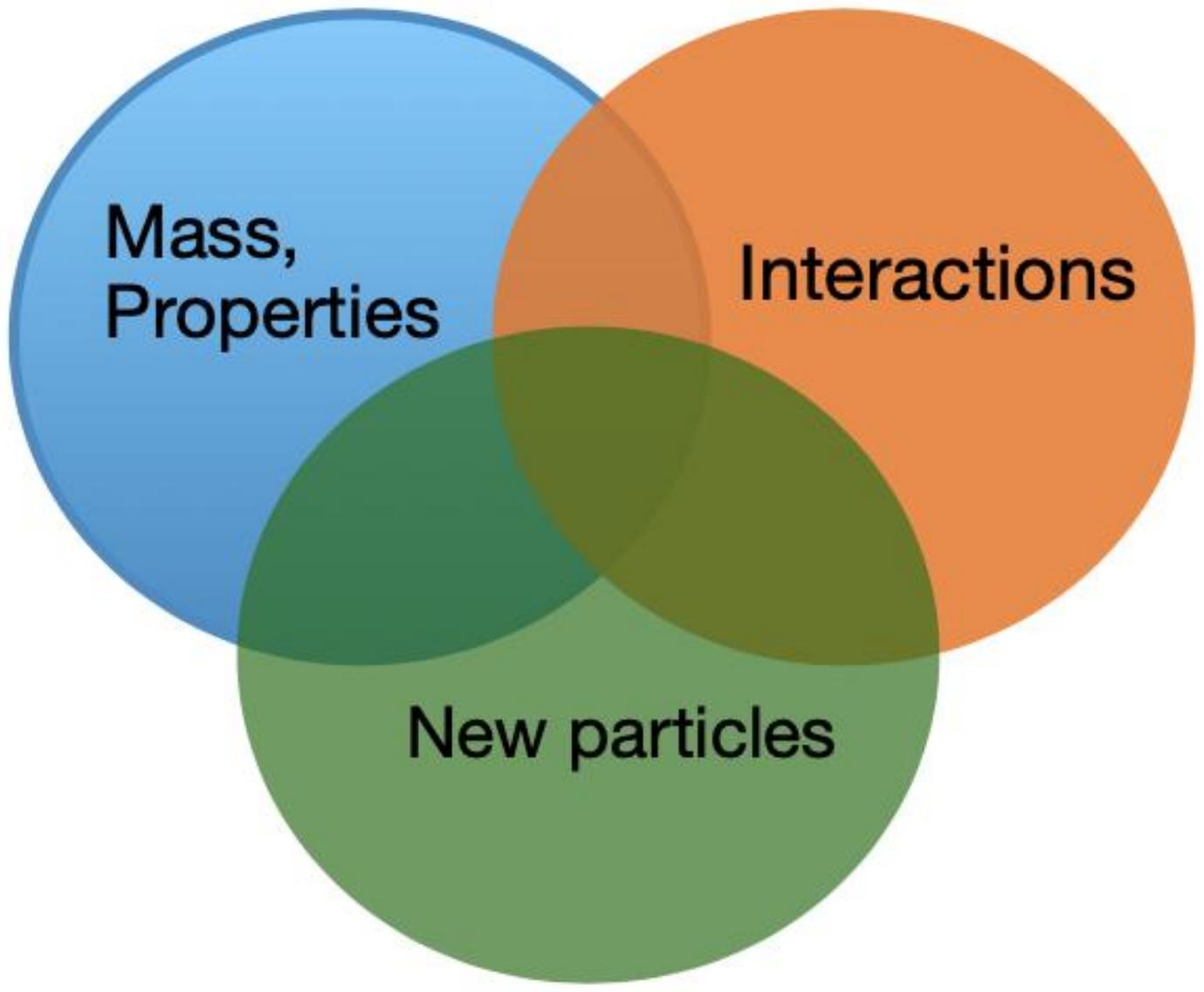}
  \caption{New physics can enter the neutrino sector through any of the above possible directions.}
 \label{fig:intro}
\end{wrapfigure}

As a result, it is crucial to probe non-standard properties of neutrinos, theoretically and experimentally, to gain insights into extensions of the SM. This holds the clue to some of the most important questions that plague particle physics today. New physics can enter the neutrino sector (i) through their mass, (ii) through their interactions, (iii) through the introduction of new particles, or (iv) through a combination of any of the above, as depicted in Fig.\,\ref{fig:intro}. 

In this article, based on the talk presented at NOW2024 , we will summarise some of the directions that can be pursued in testing non-standard neutrino physics. Some of the key directions addressed include the origin of neutrino mass and its possible connection to Dark Matter, whether neutrinos can decay, the nature of neutrinos- Dirac or Majorana or Pseudo-Dirac,  how extra species of neutrinos, called sterile neutrinos, can act as Dark Matter candidates, and new interactions of neutrinos beyond the SM. It must be emphasised that the list is by no means exhaustive and include only a subset of topics based on the personal choices of the author.

\section{Dark origin of Neutrino Mass}
The mechanism by which neutrinos acquire mass is one of the most profound questions in particle physics. The popular theoretical frameworks fall under two broad categories. Neutrinos can have a Dirac mass, which conserve lepton number. Such masses arise through couplings to a Higgs field, similar to other fermions, but require the presence of right-handed neutrino states that do not interact via SM forces. Alternatively, neutrinos, being chargeless fermions, can be Majorana particles, which will signal the violation of lepton number. Popular mechanisms such as the seesaw models~\cite{Minkowski:1977sc, Mohapatra:1979ia} can generate the small observed neutrino masses via  the Weinberg operator at dimension five.

These mechanisms rely on the Higgs or some other scalar particle obtaining a vacuum expectation value. However, it is also possible that neutrino masses have a dynamic origin, that originate due to some unknown phase transition in the past, or interaction with dark sectors~\cite{Choi:2019zxy}. We present one such scenario in~\cite{Sen:2023uga}, where neutrinos are originally massless in vacuum, but \emph{feel} an effective mass due to forward scattering on a background of ultralight dark matter (ULDM). Such ULDM, can have a uniform distribution over a large volume due to their huge de Broglie wavelength. We consider neutrinos scattering on a cold gas of ULDM $\phi$, with mass $m_\phi$, mediated by a light fermion $\chi$ with mass $m_\chi$ through the interaction $\mathcal{L}\supset g\, \bar{\chi}\, \nu\, \phi + {\rm h.c}\,$. Such a forward scattering proceeds through an $s-$channel and a $u-$ channel process, and produces a refractive potential given by~\cite{Smirnov:2021zgn,Sen:2023uga},
\begin{equation}
    V = \frac{m^2_{\rm asy}}{2 E_R}~ \frac{y - \epsilon}{y^2 - 1}\,\,\\,\,\,\,{\rm where}\,\,\,   \, m^2_{\rm asy} \equiv 
\frac{g^2\,(n_\phi + \bar{n}_\phi)}{m_\phi}\,.
\label{eq:PotPhi}
\end{equation}
Here $y=E_\nu/E_R$, with $E_R=m_\phi^2/(2m_\chi)$ being the resonant energy and $\epsilon\equiv (n_\phi - \bar{n}_\phi)/(n_\phi + \bar{n}_\phi)$ denotes the asymmetry of the DM, whose number density is given by $n$. 
\begin{wrapfigure}{r}{0.38\textwidth}
    \includegraphics[width=0.4\textwidth]{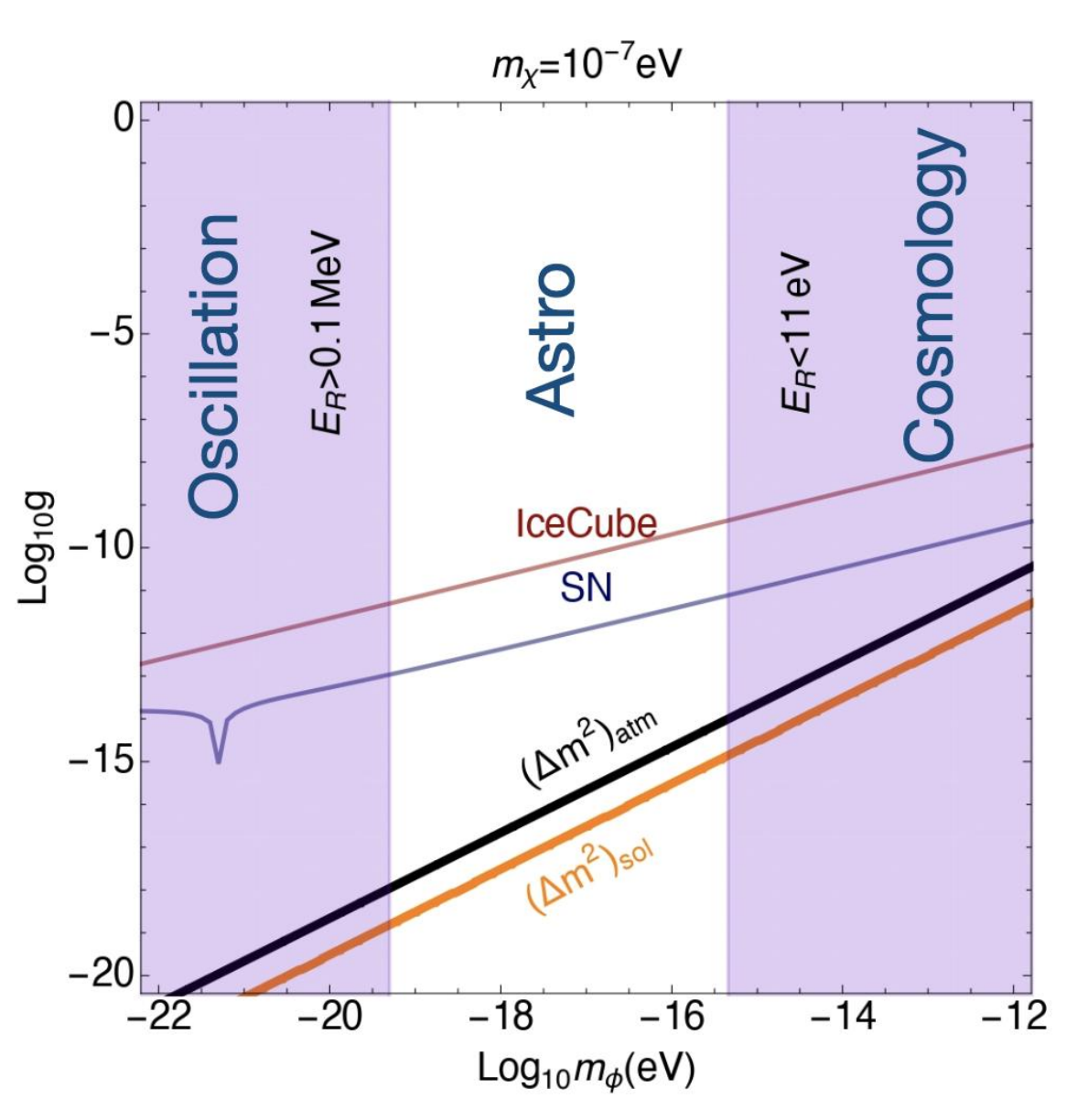}
  \caption{Allowed region. The solid black/orange lines satisfy $m_{\rm asy} ^2=\Delta m^2_{\rm atm, sol}$. The shaded regions are ruled out from oscillation experiments and cosmological bounds on $\sum m_\nu$. The region above the lines marked as ``IceCube'' and ``SN'' is ruled out~\cite{Choi:2019ixb}.  }
 \label{fig:DarkMass}
\end{wrapfigure}

The refractive mass $\tilde{m}$ can be defined through this potential as $\tilde{m}^2 \equiv 2 y E_R V  \,$ and therefore is proportional to the DM density. In the limit $y\gg 1$, note that $\tilde{m}$ goes asymptotically to $m_{\rm asy} ^2$. It presents all the properties identical to the neutrino vacuum mass-squared and hence can be used to substitute it, i.e., one can set $m_{\rm asy} ^2=\Delta m^2_{\rm atm, sol}$. As a result, in the $g-m_\phi$ parameter space, one can satisfy all oscillation data for proper choices of $m_\chi$, for example, see Fig.\,\ref{fig:DarkMass}.
On the other hand, for $y\ll 1$, we find that $\tilde{m}^2$ shows a sharp decline. Therefore, although the DM density redshifts as $(1+z)^3$, it is possible to contain $\tilde{m}^2$ within the observed limits for the sum of neutrino masses,$\sum m_\nu$,~\cite{DESI:2024mwx}, due to this sharp decline with energy. A careful analysis of the dispersion relations and the group velocities of neutrinos in such models show that neutrinos remain effectively massless for $E_R = (10 - 10^5)\,$eV~\cite{Sen:2024pgb}. In this way, the cosmological bound on the sum of the neutrino masses can be reconciled with the neutrino masses extracted from terrestrial experiments. 
\section{Non-standard decay of neutrinos and Dirac vs Majorana nature}
\begin{wrapfigure}{r}{0.45\textwidth}
    \includegraphics[width=0.45\textwidth]{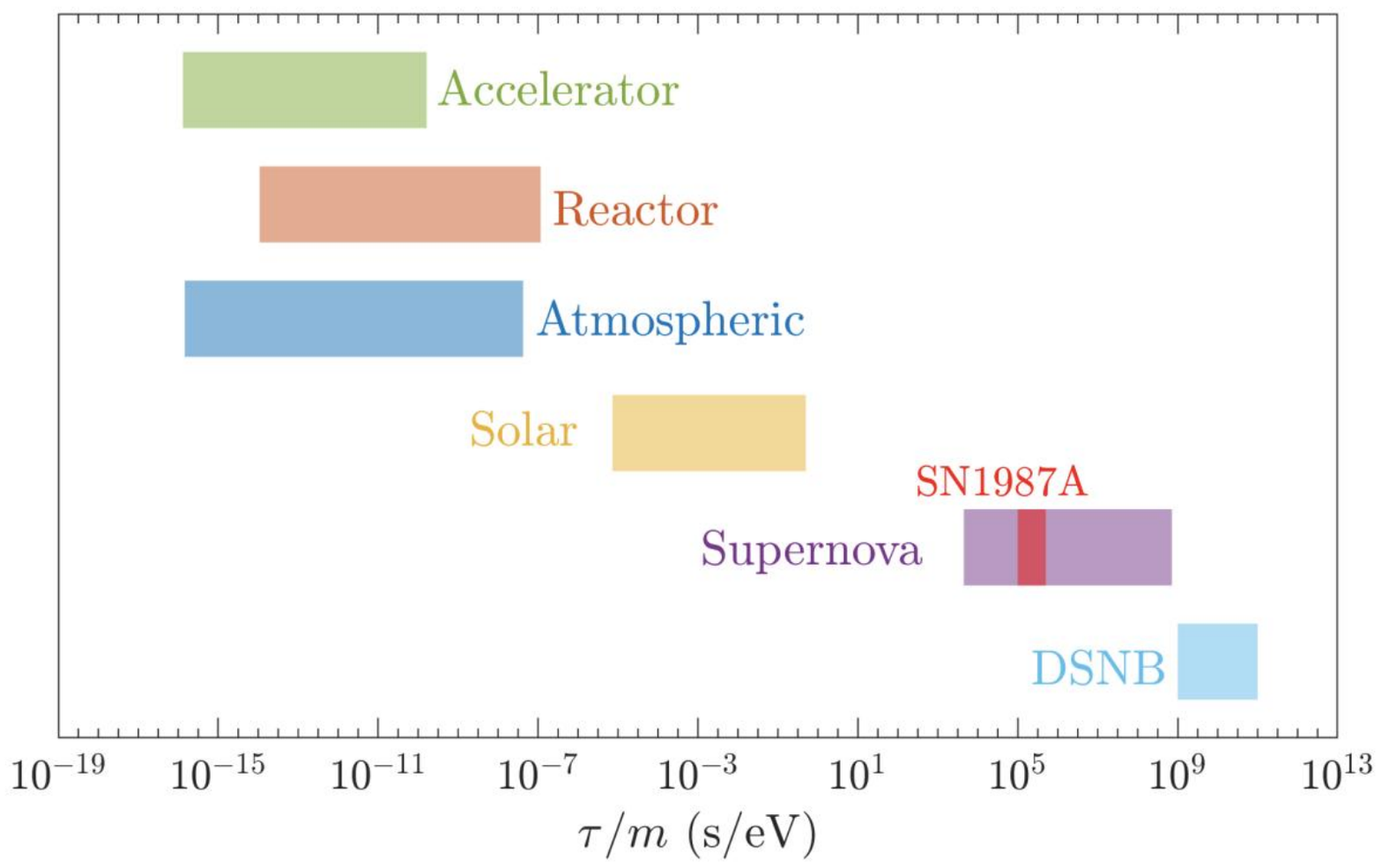}
  \caption{A summary of different constraints on $(\tau/m)$ for neutrino decay, taken from~\cite{Ivanez-Ballesteros:2023lqa}. Note that different sources probe the decay of different mass eigenstates.}
 \label{fig:Decay}
\end{wrapfigure}

Neutrinos, as massive particles, are predicted to decay into lighter states. While SM interactions predict lifetimes much longer than the age of the universe, non-standard interactions of neutrinos could mediate faster decay rates.  Such non-standard decay can proceed through new interactions like $\mathcal{L}\supset g\,\overline{\nu^c_{h}}\, \nu_{l}\, \phi$, where $\nu_h \rightarrow \nu_l $ with the emission of an almost massless scalar $\phi$. Neutrino decays can either be  visible, where the daughter products include detectable SM particles, or invisible, which involves sterile neutrinos and/or undetectable particles as final states.
Bounds on neutrino lifetime are usually quoted in terms of the ratio of lifetime over mass of the parent neutrino, $(\tau/m)$.  Since the decay is mostly sensitive to $L/E$, where $L$ is the distance travelled by neutrino, astrophysical and cosmological sources  yield the strongest bounds, while weaker constraints also exist from terrestrial experiments, see Fig.\,\ref{fig:Decay}.



%

A future galactic core-collapse SN can be used to set competitive constraints on the neutrino lifetime~\cite{deGouvea:2019goq, Ivanez-Ballesteros:2023lqa,Sen:2024fxa}. The first 25-30 ms of the neutrino burst from a SN constitutes the deleptonisation or the neutronisation burst phase~\cite{Janka:2006fh}. This phase is characterised by a huge burst in $\nu_e$, with sub-dominant contributions of $\bar{\nu}_e$ and $\nu_{\mu,\tau}$. As the neutrinos cross the Mikheyev-Smirnov-Wolfenstein resonance adiabatically, the $\nu_e$ is associated with the heaviest mass eigenstate ($\nu_3$ in Normal Ordering and $\nu_2$ in Inverted Ordering). Since the neutrino flux in this phase is almost completely $\nu_e$, this implies that in absence of decay, the $\nu_e$ received at the Earth will be around $|U_{e3}|^2$ times the original flux in Normal Ordering ($|U_{e2}|^2$ in  Inverted Ordering)~\cite{deGouvea:2019goq}. However, if the neutrino decays enroute, this fraction changes. A detector sensitive to $\nu_e$ like the upcoming Deep Underground Neutrino Experiment (DUNE) can put tight constraints on neutrino lifetime $\tau/m\sim 10^5\,{\rm s/eV}$~\cite{deGouvea:2019goq}.

Not only this, a non-standard neutrino decay can also allow one to distinguish between a decaying Dirac neutrino and a decaying Majorana neutrino~\cite{Balantekin:2018ukw, Funcke:2019grs, deGouvea:2019goq}. The idea is simple:  the decay channel mediated by the above Lagrangian has a helicity dependence, i.e., the daughter neutrino can either have the same helicity as the parent neutrino or the opposite helicity.
If neutrinos are Dirac, then the daughter neutrino with the opposite helicity behaves  as a sterile neutrino and will not show up in the detector. On the other hand, for Majorana neutrinos, this neutrino behaves like an ``antineutrino'' and will produce an $e^+$ in the detector. This can easily show up in a detector sensitive to $\bar{\nu}_e$ like Hyper Kamiokande (HK). So, if decaying neutrinos are Majorana, then HK will suddenly see a large burst of $\bar{\nu}_e$ during the neutronisation burst epoch, which will be a strong indication towards the Majorana nature of neutrinos. Thus, a combination of DUNE and HK can actually differentiate between a decaying Dirac and a decaying Majorana neutrino~\cite{deGouvea:2019goq}. 

\section{Pseudo-Dirac Neutrinos}
%
\begin{wrapfigure}{r}{0.4\textwidth}
    \includegraphics[width=0.4\textwidth]{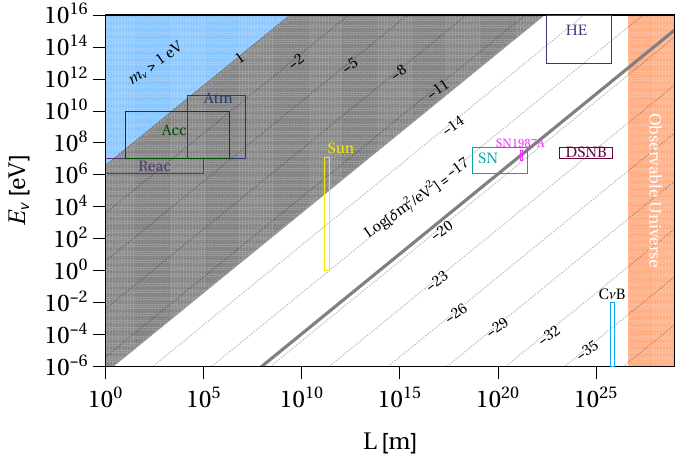}
  \caption{A list of different experiments sensitive to different $\delta m^2_{AS}$ as a function of the neutrino energy $E_\nu$ and baseline $L$~\cite{Perez-Gonzalez:2023llw}.}
 \label{fig:PD}
\end{wrapfigure}
Another intriguing scenario is when lepton number is violated softly in the SM. For neutrinos to be pseudo-Dirac, the Majorana mass must be smaller than the Dirac mass.
Diagonalisation of such a mass-matrix leads to quasi-degenerate mass eigenstates for the active and the sterile states, i.e., they have a tiny mass-squared difference, $\delta m^2_{AS}$, proportional to the smallness of the Majorana mass. 
In this case, a flavour eigenstate, $\nu_\beta$, ends up being a maximally mixed superposition of the active and sterile states, $ \nu_{\beta }=U_{\beta k}(\nu_{k}^{A}\,+\nu_{k}^{S})/\sqrt{2}$, where $U$ is the PMNS mixing matrix~\cite{Kobayashi:2000md}. Therefore, active-sterile oscillations can develop over extremely long baselines for tiny $\delta m^2_{AS}$. 

This idea can be used to constrain $\delta m^2_{AS}$ from different sources. For example, MeV neutrinos from SN1987A, coming from a distance of around $50\,{\rm kpc}$ would develop oscillations for
\begin{equation}
\delta m^2_{AS} =  \frac{4\pi E_\nu}{L_{\rm osc}}  \approx 10^{-19}{\rm eV}^2 \left(\frac{E_\nu}{20\,{\rm MeV}}\right)\left(\frac{50{\rm kpc}}{L_{\rm osc}}\right)\,.
\end{equation}
This is currently the tiniest value of $\delta m^2_{AS}$ probed so far~\cite{Martinez-Soler:2021unz}.  Other constraints arise from the observations of solar neutrinos ($\delta m^2_{AS}\lesssim 10^{-12}{\rm eV}^2$)~\cite{deGouvea:2009fp}, atmospheric neutrinos ($\delta m^2_{AS}\lesssim 10^{-4}{\rm eV}^2$)~\cite{Beacom:2003eu} as well as high energy neutrinos. Additionally, even smaller values can be probed through a measurement of the diffuse supernova background neutrinos~\cite{DeGouvea:2020ang}  or the cosmic neutrino background~\cite{Perez-Gonzalez:2023llw}. A list of current as well predicted sensitivity to values of $\delta m^2$ is listed in Fig.\,\ref{fig:PD}.

\section{Sterile Neutrinos}
Light sterile neutrinos have garnered significant attention due to their potential role in addressing anomalies in short-baseline neutrino experiments (for masses around an eV) and their viability as warm dark matter candidates (for masses around a few keV). Warm DM candidates are popular because they have been invoked to solve a number of small-scale structure problems existing between cosmological observations and N-body simulation predictions. 

\begin{wrapfigure}{r}{0.4\textwidth}
    \includegraphics[width=0.4\textwidth]{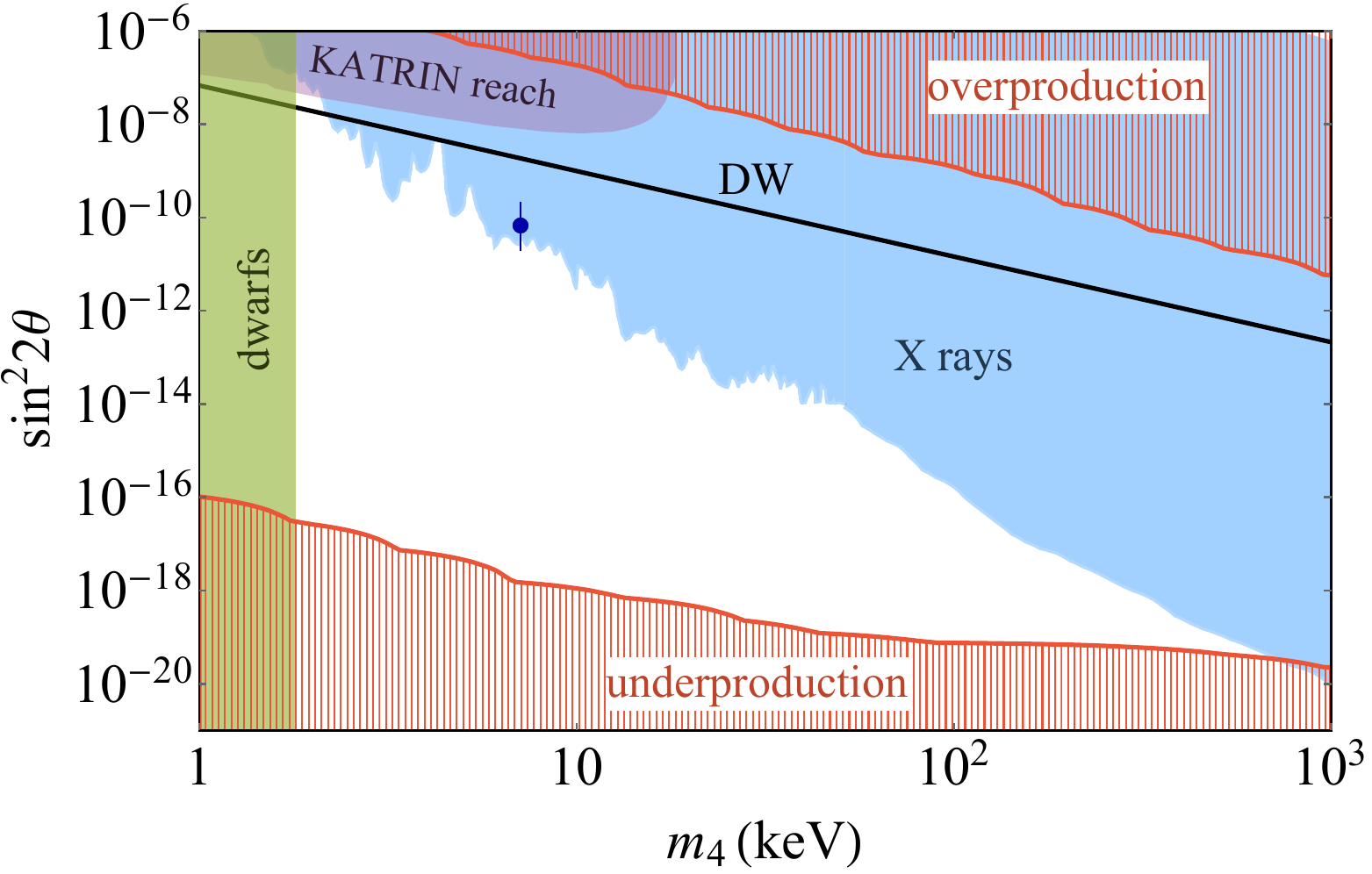}
  \caption{Constraints on sterile neutrino parameter space~\cite{DeGouvea:2019wpf}. The region between the red shaded space is allowed.}
 \label{fig:sterile}
\end{wrapfigure}

These sterile neutrinos can be produced in the early Universe through the well known Dodelson-Widrow mechanism~\cite{Dodelson:1993je}. While active neutrinos $(\nu_a)$ remain in thermal equilibrium with other SM particles, sterile neutrinos $(\nu_s)$ are out of equilibrium and assumed to have an initially negligible abundance. In the plasma, the $\nu_a$ are continually produced and propagate freely before undergoing a collision via the weak interactions.  During the propagation time, due to $\nu_a-\nu_s$ mixing, the state picks up a $\nu_s$ component. Upon a measurement  by the weak interactions, the active flavour content is reset, while the non-negligible population of $\nu_s$ is unaffected . This continues on until $\nu_a$ decouple from the cosmic plasma, leaving behind a relic population of $\nu_s$.

Currently, $\nu_s$ produced through this mechanism faces significant constraint from diffuse X-ray searches, which probe the radiative decay channel $\nu_s \rightarrow \nu_a + \gamma$. In absence of any observable unknown X-ray radiation from DM rich galaxies and halos, the sterile mass $(m_s)$ and mixing $\theta$, required to generate the observed DM relic density, is ruled out. DM masses lower than $1-2\,{\rm keV}$ also faces strong bound from phase-space considerations derived for dwarf galaxies, as well as Lyman-alpha forest data~\cite{Abazajian:2017tcc}.  

A popular and testable way to efficiently produce $\nu_s$ DM without being ruled out by these constraints is to postulate non-standard self-interactions of active neutrinos~\cite{DeGouvea:2019wpf}. This can arise due to an interaction of the form, $\mathcal{L} 	\supset \lambda_\phi\, \overline{\nu_L^c}\, \nu_L \,\phi + {\rm h.c.} \,$, where $\phi$ is a neutrinophilic scalar with mass $m_\phi$. There is nothing special about a scalar interaction, and one can present the entire argument with vector mediators as well.  These non-standard interactions can be constrained from invisible Higgs decays, Z-boson decays, tau decays, decays of heavy mesons, neutrinoless double-beta decays, accelerator neutrino experiments as well as cosmological surveys, however, in general, the constraints are quite weak~\cite{Blinov:2019gcj}.
Typically, these interactions can be much stronger than the electroweak interactions. Since the rate of these interactions is much larger, it allows the relic density of DM to be produced at smaller mixing angles, thereby alleviating the tension with the X-ray bounds~\cite{DeGouvea:2019wpf}. This extends the allowed parameter space from a narrow line to a broad band (see Fig.\,\ref{fig:sterile}) and helps in reviving the sterile neutrino dark matter candidates. 

\section{Non-standard neutrino interactions}
As alluded to in the previous sections, neutrinos can have interactions beyond the SM. Such non-standard interactions (NSI) can happen between neutrinos and quarks/leptons, or in the form of neutrino non-standard self-interactions (NSSI). These are some of the most promising directions to look for new physics in the neutrino sector. There is a vast literature discussing the constraints on neutrino NSI as well as NSSI, please refer to \cite{Proceedings:2019qno,Berryman:2022hds} and the references therein for a detailed discussion. 

A particular subclass of such non-standard interaction is neutrino-dark matter interactions. A popular way to probe neutrino-DM interactions  is through its impact on the cosmic microwave background, or through DM indirect detection tests. Another novel direction is through the detection of boosted DM. The DM in the Milky Way halo can be upscattered through interactions with high energy astrophysical neutrinos and can leave unique signatures in direct detection experiments. For example, neutrinos from the diffuse supernova neutrino background can boost a fraction of the DM to relativistic (MeV) energies, which can be easily detected in experiments like XENONnT, LZ and PandaX~\cite{Das:2024ghw}. On the other hand, absence of such a signature can allow one to constrain a combination of DM-neutrino and DM-electron constraints. Similar studies have also been performed for solar neutrinos, SN neutrinos as well as cosmic ray electrons and protons.

\section{Concluding remarks}
Neutrinos remain an unparalleled tool for investigating new physics. Their unique properties and interactions provide insights into fundamental questions about mass, their nature, as well as relation, if any, with dark matter.  This presentation highlights the interplay between neutrino properties and broader questions within the Standard Model, paving the way for future discoveries. Continued experimental efforts, coupled with theoretical advancements, will undoubtedly yield transformative discoveries in our understanding of the universe.

\section*{Acknowledgements}
I thank the organizers of NOW2024 for hosting an amazing workshop. I would also like to thank all colleagues who have collaborated with me in the projects that contributed to this talk.

\end{document}